\documentclass[fleqn,usenatbib]{mnras}

\usepackage{newtxtext,newtxmath}

\usepackage[T1]{fontenc}
\usepackage{ae,aecompl}

\usepackage{graphicx}	
\usepackage{amsmath}	
\usepackage{amssymb}	
\usepackage{mathtools}
\usepackage{subfig}
\usepackage{caption}
\usepackage{cleveref}
\crefformat{section}{\S#2#1#3} 
\crefformat{subsection}{\S#2#1#3}
\crefformat{subsubsection}{\S#2#1#3}

\title[ Probing the high-z IGM with $^3$He$^+$ signal]{ Probing the high-z IGM with the hyperfine transition of $^3$He$^+$}

\author[S. Khullar et. al.]{Shivan Khullar$^{1,2,3},$\thanks{E-mail: shivankhullar@gmail.com (SK)}
Qingbo Ma$^{4}$,
Philipp Busch$^{5,6,1}$,
Benedetta Ciardi$^{1}$, \newauthor
Marius B. Eide$^{1}$,
Koki Kakiichi$^{7}$
\\
$^{1}$Max-Planck-Institut f\"ur Astrophysik, Karl-Schwarzschild-Str. 1, 85748 Garching, Germany\\
$^{2}$Research School of Astronomy and Astrophysics, Australian National University, Canberra, ACT~2611, Australia\\
$^{3}$Department of Physics, Goa Campus, Birla Institute of Technology and Science, Pilani, Rajasthan, 333031, India\\
$^{4}$Guizhou Provincial Key Laboratory of Radio Astronomy and Data Processing, Guizhou Normal University, Guiyang 550001, PR China\\
$^{5}$Department of Natural Science, The Open University of Israel, 1 University Road, P. O. Box 808, Raanana 43107, Israel\\
$^{6}$Department of Physics, The Technion, Haifa 3200003, Israel\\
$^{7}$Department of Physics and Astronomy, University College London, London, WC1E 6BT, UK
}

\date{Accepted XXX. Received YYY; in original form ZZZ}

\pubyear{2020}

\begin{document}
\label{firstpage}
\pagerange{\pageref{firstpage}--\pageref{lastpage}}
\maketitle


\begin{abstract}

The hyperfine transition of $^3$He$^+$ at 3.5~cm has been thought as a probe of the high-$z$ IGM since it offers a unique insight into the evolution of the helium component of the gas,  as well as potentially give an independent constraint on the 21~cm signal from neutral hydrogen. In this paper, we use radiative transfer simulations of reionization driven by sources such as stars, X-ray binaries, accreting black holes and shock heated interstellar medium, and simulations of a high-$z$ quasar to characterize the signal and analyze its prospects of detection. We find that the peak of the signal lies in the range $\sim 1-50$~$\mu$K for both environments, but while around the quasar it is always in emission, in the case of cosmic reionization a brief period of absorption is expected. As the evolution of HeII is determined by stars, we find that it is not possible to distinguish reionization histories driven by more energetic sources. On the other hand, while a bright QSO produces a signal in 21~cm that is very similar to the one from a large collection of galaxies, its signature in 3.5~cm is very peculiar and could be a powerful probe to identify the presence of the QSO. We analyze the prospects of the signal's detectability using SKA1-mid as our reference telescope. We find that the noise power spectrum dominates over the power spectrum of the signal, although a modest S/N ratio can be obtained when the wavenumber bin width and the survey volume are sufficiently large.

\end{abstract}




\section{Introduction}
\label{sec:Intro}

The reionization of hydrogen and helium in the inter-galactic medium (IGM) at redshifts $z \gtrsim 6$ and $z \gtrsim 2.5$, respectively, is the subject of continued investigations (for a review see \citealt{Ciardi.Ferrara_2005, MoralesWyithe2010, PritchardLoeb2012,  LoebFurlanetto2013}). Helium reionization is considered to be the last major phase change in the Universe. While the radiation from stellar type sources  responsible for hydrogen reionization singly ionize helium as well,  photons with higher energies are required to ionize helium fully. In fact, theoretical models (e.g. \citealt{Compostella.Cantalupo.Porciani_2014,LaPlante.Trac_2016}), as well as observations (e.g. \citealt{Worseck_etal_2011}), indicate that helium reionization is driven by Quasi Stellar Objects (or quasars; QSOs).

While QSOs' spectra are routinely used to probe the final phases of hydrogen (e.g.  \citealt{Becker.Bolton.Lidz_2015})  as well as helium (e.g. \citealt{Worseck.Prochaska.Hennawi.McQuinn_2016}) reionization, little observational constraint is available on their history.
The hyperfine transition of neutral hydrogen (HI), with a rest frame frequency of 1.42 GHz (21~cm), is a promising probe of the evolution of H reionization (for a review see \citealt{Furlanetto2006}). Several ongoing observational efforts have been made in this respect. Among these, the LOw Frequency ARray\footnote{www.lofar.org} (LOFAR), the Murchison Widefield Array\footnote{http://www.mwatelescope.org} (MWA),  the Hydrogen Epoch of Reionization Array\footnote{https://reionization.org} (HERA), the Precision Array for Probing the Epoch of Reionization\footnote{http://eor.berkeley.edu} (PAPER) and the upcoming Square Kilometer Array\footnote{https://www.skatelescope.org} (SKA). 
While efforts to probe the high-$z$ IGM have focused extensively on the 21~cm line, the hyperfine transition of $^3$He$^+$ at 8.66~GHz (3.5~cm) also offers a unique insight into some astrophysical phenomena prevalent in the early universe \citep{Furlanetto2006,BaglaLoeb2009,McQuinn.Switzer_2009,Takeuchi.Zaroubi.Sugiyama_2014,Vasiliev2019}. In addition to being a probe of the high-$z$ evolution of the helium component of gas, \citet{BaglaLoeb2009} suggested that this feature could also in principle be used to obtain independent constraints on the 21~cm signal, because the evolution of both HII and HeII are driven by stellar type sources during the epoch of (hydrogen) reionization, and thus the 21~cm and 3.5~cm signals are expected to be anti-correlated at these redshifts. 
Finally, in a QSO-dominated reionization scenario (e.g. \citealt{Madau.Haardt_2015,Hassan_etal_2018}), the abundance of HeII is much lower than in a standard model because of the hard spectrum of QSOs. This means that, in principle, the strength of the 3.5~cm signal can constrain the type of sources dominating the reionization process, and potentially even their relative importance.

Although the abundance of HeII is much smaller than that of HII, the above authors have suggested that the $^3$He$^+$ hyperfine transition offers some advantages over the 21~cm line, which are summarized as follows: {\it (i)} as its rest frame frequency is considerably higher than the corresponding frequency for H, observations of this signal would suffer from less severe foreground contamination; and {\it (ii)} the spontaneous decay rate of the $^3$He$^+$ transition is $\sim$ 680 times larger, thus boosting the signal. However, the prospects for the detection of this signal are severely limited with current telescopes. Although several radio telescopes are operational in the relevant frequency range, probably the best chance to detect the 3.5~cm signal lies with SKA. 

Although \citet{BaglaLoeb2009} have evaluated the expected signal with a semi-analytic approach, a more rigorous modeling has yet to be done which accounts for the evolution of HeII (the thorough analysis of \citealt{Takeuchi.Zaroubi.Sugiyama_2014} was concentrated at lower redshift and still lacked a full radiative transfer). Here we revisit this problem employing the simulations of reionization described in \citet[hereafter E18]{Eide_etal_2018} and Eide et al. (submitted; hereafter E20), which model both H and He reionization as driven by stellar type sources, accreting nuclear black holes, X-ray binaries and shock heated interstellar medium. We will repeat the same analysis concentrating on the enviroment surrounding the high-$z$ quasar described in \citet[hereafter K17]{Kakiichi2017}. 

The rest of the paper is structured as follows. In \cref{sec:Simulations and Method} we describe the simulations presented in E18, E20 and K17 and the methodology used to calculate the $^3$He$^+$ signal from such simulations. In \cref{sec:Results} we present the results of our analysis and we summarize our findings, and highlight the main conclusions of our study in \cref{sec:Discussion and Conclusions}.

\section{Simulations and Method}
\label{sec:Simulations and Method}

In this work we will evaluate the signal associated to the hyperfine transition of the $^3$He$^+$ both on cosmological scales and around a high-$z$ QSO.
The simulations used here are those described in E18 and E20 for the cosmological signal, and those discussed in K17 for the QSO environment. Here we outline their main characteristics and refer the readers to the original papers for more details. 

\subsection{Simulations of cosmic reionization}
\label{sec:simul}

Outputs of the MassiveBlack-II (MBII; \citealt{Khandai_etal_2015}), a high resolution cosmological SPH simulation, are combined with population synthesis modeling of ionizing sources and post-processed with the multi-frequency 3D radiative transfer code {\tt CRASH} (e.g. \citealt{Ciardi2001}, \citealt{Maselli2003}, \citealt{Maselli2009}, \citealt{Graziani2018}) to model hydrogen and helium reionization. 
The MBII simulation has been run 
using P-GADGET (see \citealt{Springel2005b} for an earlier version of the code) and tracks stellar populations, galaxies, accreting and dormant black holes as well as their properties like age, star formation rate, metallicity, mass, accretion rate, etc.  The simulation is performed in the WMAP7 $\Lambda$CDM cosmology \citep{Komatsu2011}, has a box length of $100h^{-1}$ cMpc and $2\times 1792^3$ gas and dark matter particles, with a mass of m\textsubscript{gas} $=  2.2 \times 10^6 h^{-1}$M\textsubscript{\(\odot\)} and m\textsubscript{DM} = $1.1 \times 10^7 h^{-1}$ M\textsubscript{\(\odot\)}, respectively. For outputs with redsihft in the range $z=6-20$ the gas, temperature and ionization fractions fields, as well as the sources of ionizing photons, have been mapped onto $ N=256^3$ grids to be post-processed with {\tt CRASH}, assuming an escape fraction of UV photons (13.6~eV$< h\nu <$ 200~eV) of 15\%. The source types included in the radiative transfer simulation are stars, X-Ray Binaries (XRBs), accreting nuclear Black Holes (BHs) and bremsstrahlung from shock heated interstellar medium (ISM). For more information on the sources and their effects on the IGM, we refer the reader to E18 and E20. 

\subsection{Simulations of high-$z$ QSO}
\label{sec:QSO}

Similarly to the model of reionization described above, a
hydrodynamical simulation of the IGM run with GADGET-3 has been post-processed with {\tt CRASH} to investigate the environment surrounding a high-$z$ QSO. The simulation adopts cosmological parameters consistent with WMAP9 results \citep{Hinshaw_et_al2013}, has a box length of $50h^{-1}$ cMpc and contains $2\times 512^3$ gas and dark matter particles, corresponding to a mass of m\textsubscript{gas} $= 1.2 \times 10^7 h^{-1}$M\textsubscript{\(\odot\)} and m\textsubscript{DM} $= 5.53 \times 10^7 h^{-1}$M\textsubscript{\(\odot\)}, respectively. By design, the simulation box is centered on the largest halo, having a mass of $1.34 \times 10^{10}h^{-1}$M\textsubscript{\(\odot\)} at $z=10$. For outputs in the redshift range $z = 15-10$, the gas and temperature fields are mapped onto grids with $N=256^3$ cells and fed as input to {\tt CRASH} to solve for the radiative transfer of UV photons emitted by the stellar sources. At $z=10$ we assume that a QSO turns on in the center of the most massive halo and its much harder radiation (in this case we follow the RT also in the soft X-ray regime, i.e. 200~eV$< h\nu <$ 2~keV) is evolved for a time corresponding to the lifetime of the QSO. For further details about the source model and its impact on the ionization and thermal state of the QSO environment, we refer the reader to K17. 

\subsection{$^3$He$^+$ signal}
\label{sec:DBT}

The simulations of cosmic reionization and QSO's environment described in the previous sections provide, among others, the spatial and temporal distribution of gas temperature, $T_{\rm gas}$, as well as fractions of HII, HeII and HeIII (i.e. $x_{\rm HII}$, $x_{\rm HeII}$, and $x_{\rm HeIII}$, respectively). The gas number density, $n_{\rm gas}$, is instead taken directly from the hydrodynamic simulations. From these quantities, the differential brightness temperature associated with the hyperfine transition of $^3$He$^+$, $\delta T_{\rm b,^3He}$, can be evaluated in each cell of the simulated volume as (see eq. 61 of \citealt{Furlanetto2006}):

\begin{multline}
    \label{eq:DBT-Field}
\delta T_{\rm b,^3He} \approx 0.5106 \ x_{\rm HeII} (1+\delta) \left( 1 - \frac{T_{\rm CMB}}{T_{\rm s}}\right)\\ \ \left( \frac{\rm [^{3}He/H]}{10^{-5}} \right) \left( \frac{\Omega_b h^2}{0.0223}\right) \sqrt{\frac{\Omega_m}{0.24}}  (1+z)^{1/2}  {\rm \mu K},
\end{multline}

where $\delta = (n_{\rm gas} - \bar{n}_{\rm gas})/\bar{n}_{\rm gas}$ is the gas overdensity with $\bar{n}_{\rm gas}$ mean gas number density of the whole box, $x_{\rm HeII}$ is the fraction of HeII, $T_{\rm CMB} = 2.725 (1+z)$~K is the CMB temperature at redshift $z$, $T_{\rm s}$ is the spin temperature, $\Omega_b$ and $\Omega_m$ are the baryonic and matter density parameters and $h = H_0/100$, where $H_0$ (in units of km~s$^{-1}$~Mpc$^{-1}$) is the Hubble constant. The abundance of $^{3}$He relative to that of H, $[^{3}$He/H]$\approx 10^{-5}$, is dictated by the big bang nucleosynthesis (eq.~6 of \citealt{A_Coc}).  Note that in the equation above we have made the common assumption that $\frac{H(z)}{(1+z)(dv_{\parallel}/dr_{\parallel})} \sim 1$, and throughout the paper we will also assume that $T_{\rm s} \sim T_{\rm gas}$ (see Appendix~\ref{app} for a discussion on the latter assumption).

As the power spectrum (PS) is a quantity which can be inferred directly from radio interferometric observations, we will also estimate the PS of the $^3$He$^+$ signal as expected from our simulations. We evaluate the PS of the differential brightness temperature field as: 
\begin{equation}
\label{eq:Pk}
P({\bf k}) = \langle \delta T_{\rm b,^3He} ({\bf k}) \  \delta T_{\rm b,^3He}({\bf k})^* \rangle,
\end{equation}

where $\delta T_{\rm b,^3He} ({\bf k})$ is the Fourier transform of the differential brightness temperature field, and $\delta T_{\rm b,^3He}({\bf k})^*$ is the complex conjugate of $\delta T_{\rm b,^3He} ({\bf k})$. We additionally define the quantity $\Delta^2_{\rm ^3 He}$ (in units of [$\mu$ K]$^2$) as:

\begin{equation}
\label{eq:Deltasquare}
\Delta^2_{\rm ^3 He} = \frac{k^3}{2\pi^2} P({\bf k}).
\end{equation}

\section{Results}
\label{sec:Results}

In this section, we will briefly discuss the properties of the output fields of the simulations mentioned in \cref{sec:Simulations and Method}, the differential brightness temperature field and the power spectrum of the $^3$He$^+$ signal at different redshifts. 

\subsection{Cosmic reionization}
\label{sec:reion} 


While we refer the reader to E18 and E20 for an extensive discussion of the evolution of the IGM temperature and ionization state, here we briefly highlight what is strictly necessary for this study.

\begin{figure*}
	\includegraphics[width=\textwidth]{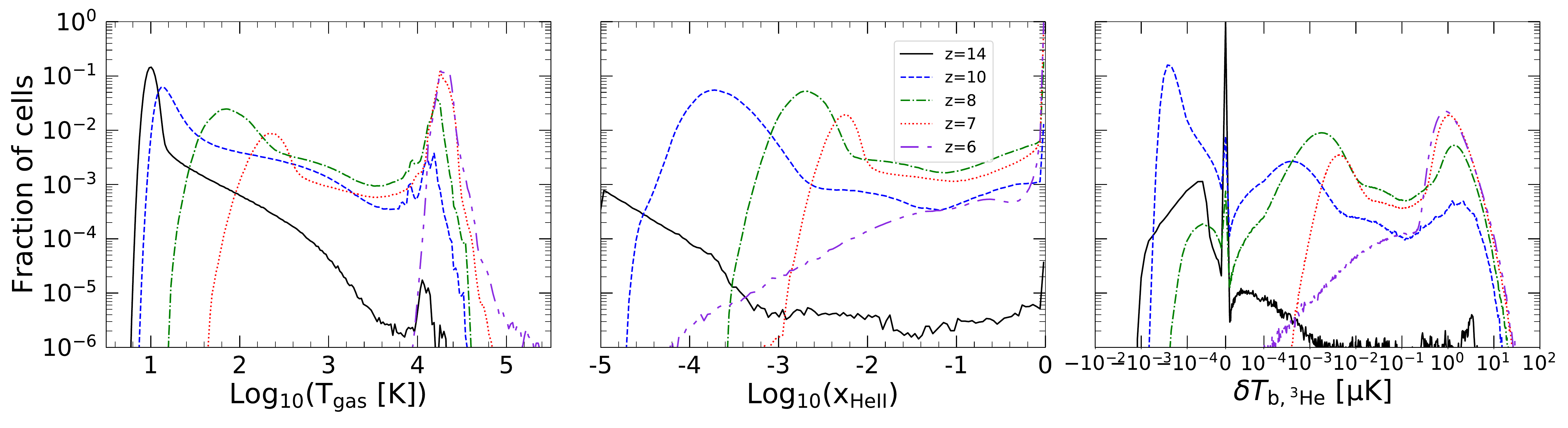}
   \caption{From left to right the panels refer to the distribution of the gas temperature, $T_{\rm gas}$, of the HeII fraction, $x_{\rm HeII}$, and of the differential brightness temperature, $\delta T_{\rm b,^3He}$, for the 256$^3$ cells in the cosmological simulation of cosmic reionization as described in \cref{sec:simul}. The lines refer to $z=14$ (black solid), 10 (blue dashed), 8 (green dash-dotted), 7 (red dotted) and 6 (purple dash-dot-dotted). }
    \label{fig:PDF-GXQI}
\end{figure*}

In Figure \ref{fig:PDF-GXQI} (left panel) we show the distribution of the gas temperature. At high redshift it presents two peaks: the strongest one is at $T_{\rm gas} \sim 10$~K \footnote{Note that this reflects the numerical temperature floor of the simulations.}, indicating that most of the gas is still cold, while a second considerably weaker peak is visible at $T_{\rm gas} \sim 10^4$~K, corresponding to the gas that has already been ionized by the few sources present at these redshifts.  As the redshift decreases and more, larger sources appear increasing the size and number of ionized regions, the first peak shifts towards larger temperatures and becomes less relevant, until it completely disappears by $z \sim 6$ when all gas is ionized. At the same time, the second peak becomes predominant and shifts as well towards larger values as the helium component of the gas gets ionized in appreciable quantities.  
 
Since the first ionization potential of He at $24.6 $ eV is close to the $13.6$ eV for H, all H ionizing sources in the simulation singly ionize He as well, and as a consequence the distribution of $x_{\rm HeII}$ and $x_{\rm HII}$ resemble each other very closely, except at low redshift when an appreciable fraction of $x_{\rm HeIII}$ (the second ionization potential of He is 54.4~eV) starts to appear\footnote{For more information on the sources' spectra and their effect on the physical state of the IGM we refer the reader to E18 and E20.}.
For this reason, in Figure \ref{fig:PDF-GXQI} (central panel) we show only the distribution of $x_{\rm HeII}$, which, similarly to the gas temperature, presents two peaks, one at low ionization which shifts towards larger values with decreasing redshift (and eventually disappears), and one corresponding to full ionization, which becomes increasingly important as reionization proceeds\footnote{Note that the Monte Carlo method adopted here assures convergence of the results down to a value of the ionization fraction of $10^{-5}$.}. 

\begin{figure}
	\includegraphics[width=\columnwidth]{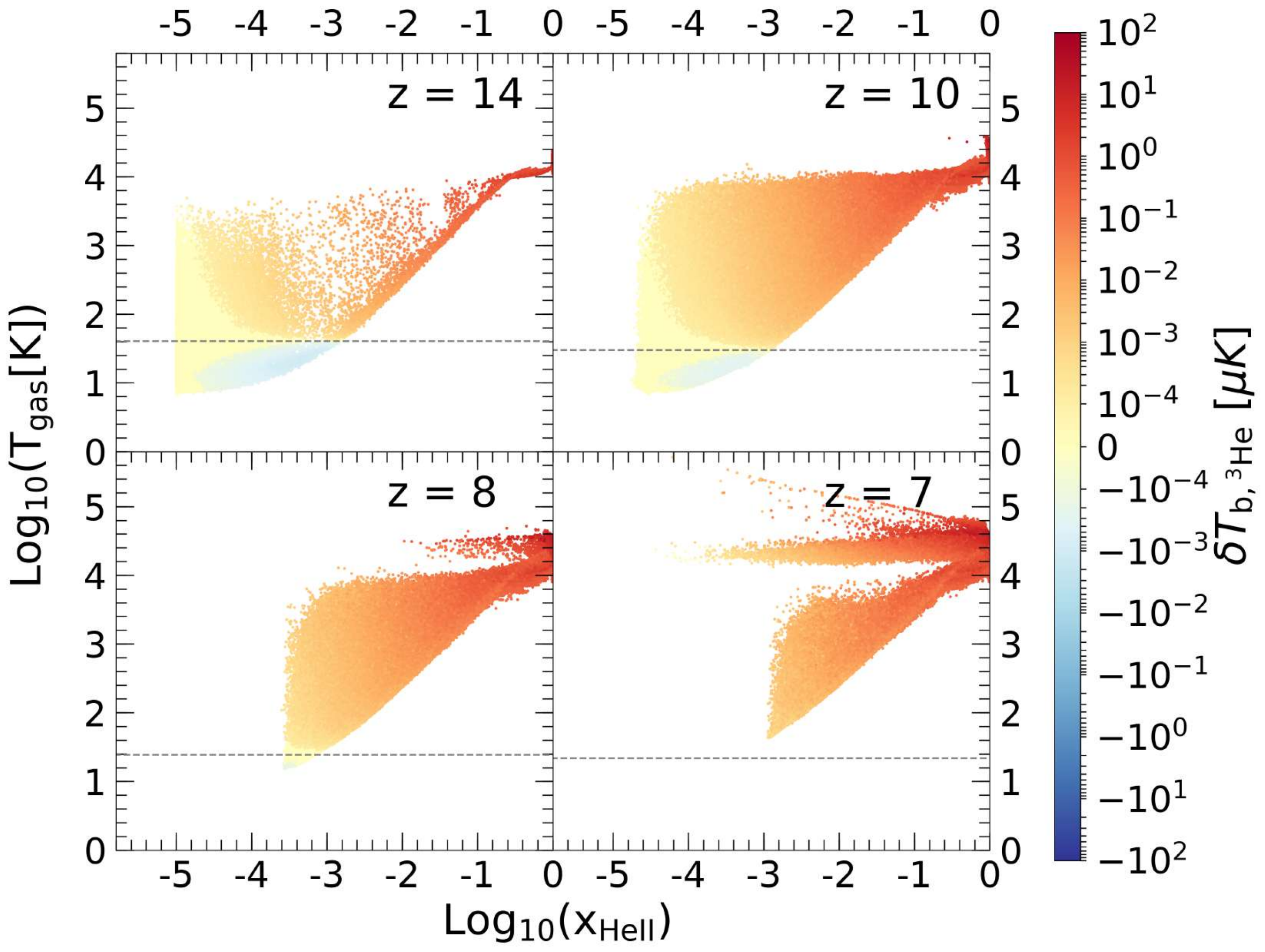}
   \caption{Differential brightness temperature, $\delta T_{\rm b,^3He}$, for the 256$^3$ cells in the cosmological simulation of cosmic reionization as described in \cref{sec:simul} as a function of the corresponding $T_{\rm gas}$ and $x_{\rm HeII}$ values. The panels refer to $z=14$ (top left), 10 (top right), 8 (bottom left) and 7 (bottom right). The horizontal lines indicate $T_{\rm CMB}$.}
    \label{fig:DBT-Temp-xHeII}
\end{figure}

The corresponding distribution of the differential brightness temperature field is shown in the right panel of Figure \ref{fig:PDF-GXQI}. Due to the small amount of singly ionized He, at high redshifts $\delta T_{\rm b,^3He}$ is mostly  zero. Nevertheless, the signal can be observed in both absorption and emission with values varying between $\sim -10^{-3}$ and $\sim 5$~$\mu$K. 
The negative values are associated to partially ionized cells whose temperature has not been raised above that of the CMB (see also Fig.~\ref{fig:DBT-Temp-xHeII}) and which contribute to the cold peak observed in the left panel of the  Figure. The peak in the negative values grows from $z=14$ to $z=10$ since there are more partially ionized cells at $z=10$ than $z=14$, as is also seen in the central panel of Figure \ref{fig:PDF-GXQI}. As the redshift decreases, more helium gets ionized and the gas heated, causing the cells with negative and zero values of $\delta T_{\rm b,^3He}$ to diminish and eventually disappear when reionization is complete by $z \sim 6$. At the same time, most of the signal becomes in emission and increases in intensity. 

For a better understanding of the dependence of the differential brightness temperature on $T_{\rm gas}$ and $x_{\rm HeII}$, in Figure~\ref{fig:DBT-Temp-xHeII} 
we plot a phase diagram at various redshifts. The value of the differential brightness temperature for each of the $256^3$ cells has been plotted as a function of the corresponding $x_{\rm HeII}$
and $T_{\rm gas}$ values. At redshifts $z \geq 10$, the main contribution to the differential brightness temperature comes, at negative values, from gas which is colder than $T_{\rm CMB}$ and poorly ionized 
($x_{\rm HeII} \le 10^{-3}$), and, at positive values, from highly ionized hot gas. As $z$ decreases, the gas transitions from the cold to the hot peak that was seen in Figure \ref{fig:PDF-GXQI} while $x_{\rm HeII}$ increases, resulting in a signal which is predominantly (and eventually totally) in emission. Once $T_{\rm gas} \gg 10^4$~K, helium becomes doubly ionized and consequently $x_{\rm HeII}$ decreases. 

\begin{figure}
	\includegraphics[width=\columnwidth]{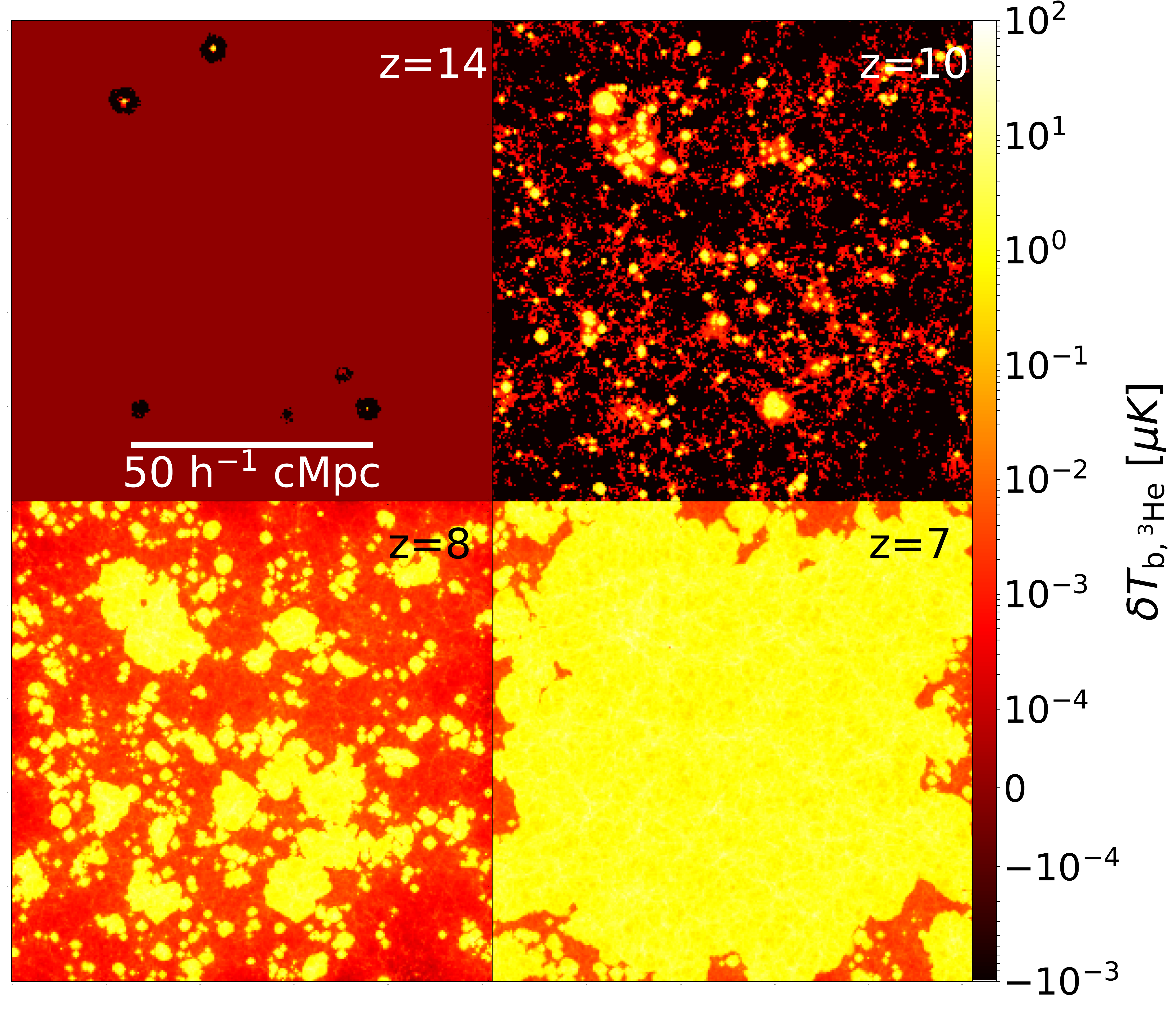}
   	\caption{A slice of the differential brightness temperature field (in $\mu$K) in the cosmological simulation of cosmic reionization as described in \cref{sec:simul} at $z=14$ (top left panel), 10 (top right), 8 (bottom left) and 7 (bottom right).}
    \label{fig:DBT-Field}
\end{figure}

In Figure \ref{fig:DBT-Field} we show maps of $\delta T_{\rm b,^3He}$ in a slice of the simulation box. At redshift $z \geq 10 $ the IGM is either predominantly neutral or very lowly ionized (as a result of hard ionizing photons with a large mean free path like the ones emanating from energetic sources such as XRBs, ISM and BHs) and thus in most of the IGM $\delta T_{\rm b,^3He}$ is zero or in the range $[-10^{-4},-10^{-3}] \ \mu $K.
As the redshift decreases, the differential brightness temperature values increase and reach a maximum of $\delta T_{\rm b,^3He} \sim 25 \ \mu $K; the amount of HeII on large scales increases (as the ionized regions grow in size and number) and the $^3$He$^+$ signal can only be seen in emission. 

\begin{figure}
	\includegraphics[width=\columnwidth]{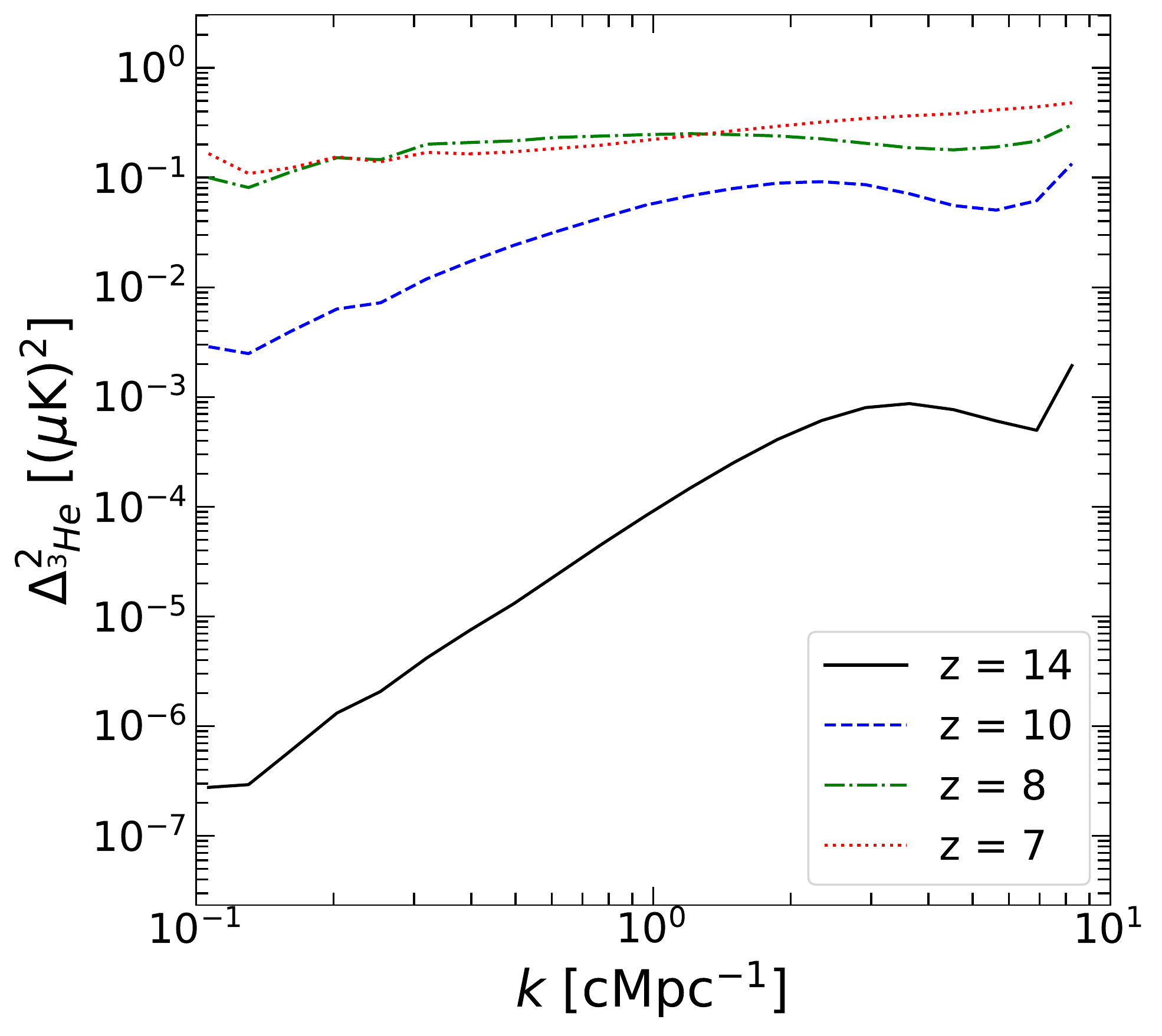}
   \caption{Power spectrum of the $^3$He$^+$ signal for the simulation of cosmic reionization as described in \cref{sec:simul} at redshift $z=14$ (black solid line), 10 (blue dashed line), 8 (green dash-dotted line), and 7 (red dotted line). }
    \label{fig:PS-He-GXQI}
\end{figure}

In Figure \ref{fig:PS-He-GXQI} we show $\Delta^2_{\rm ^3 He}$ as a function of the wavenumber $k$. At high redshifts, the intensity of the power spectrum is extremely low since there is very little HeII in the simulation box and, as it is concentrated in small ionized regions around the early galaxies, $\Delta^2_{\rm ^3 He}$ behaves like white gaussian noise. However, as reionization proceeds and more HeII gets ionized, we observe a general increase of power, in particular at small $k$ values. As a result, the shape of the power spectrum becomes flatter with decreasing redshift, until the power is almost the same on all scales at $z<7$ when $x_{\rm HeII} \sim 1$ everywhere. Although not shown here, we expect the shape of the HeII 3.5~cm signal to be similar to that of the HII 21~cm signal, as the morphology of the HII and HeII bubbles resemble each other in the absence of strong 54.4 eV ionizing sources.  

\subsection{High-$z$ QSO}
\label{sec:QSOhz} 

\begin{figure*}
	\includegraphics[width=\textwidth]{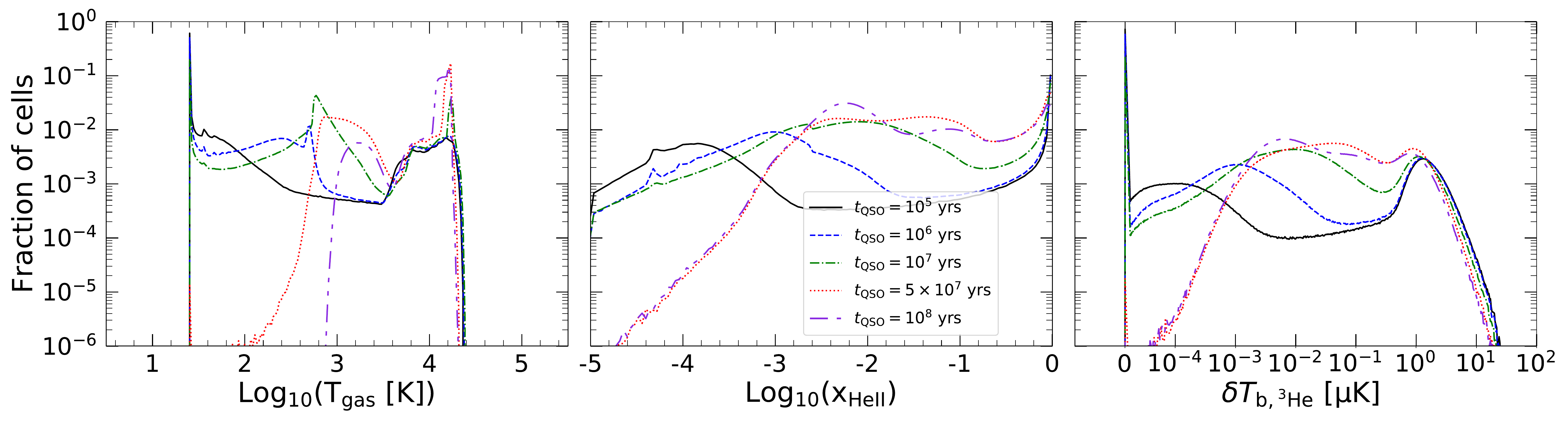}
   \caption{From left to right the panels refer to the distribution of the gas temperature, $T_{\rm gas}$, of the HeII fraction, $x_{\rm HeII}$, and of the differential brightness temperature, $\delta T_{\rm b,^3He}$, for the 256$^3$ cells in the simulation of a high-$z$ QSO as described in section~\cref{sec:QSO}. The lines refer to a quasar's lifetime of $t_{\rm QSO}=10^5$~yrs (black solid), $10^6$~yrs (blue dashed), $10^7$~yrs (green dash-dotted), $5 \times 10^7$~yrs (red dotted) and $10^8$~yrs (purple dash-dot-dotted). }
    \label{fig:PDF-QSO}

\end{figure*}

As in the case of the simulations of cosmic reionization, we refer the reader to K17 for an extensive discussion of the QSO's environment and its impact on the ionization and thermal state of the IGM, while we only briefly highlight what is strictly necessary for this study. In Figure \ref{fig:PDF-QSO} (left panel) we show the distribution of the gas temperature for different values of the QSO's lifetime, $t_{\rm QSO}$. Similarly to what noted in the previous section, initially two clear peaks are observable, one associated to the cold gas that has not yet been reached by ionizing photons, and the second at $T_{\rm gas} \sim 12500$~K, accounting for the fully ionized gas surrounding the central QSO and/or nearby galaxies. As expected, the first (second) peak becomes less (more) prominent for larger $t_{\rm QSO}$, as an increasing number of cells gets fully ionized. A peak at intermediate values is also visible, which corresponds to the temperature of the central HeIII bubble within the HeII region. The peak at the highest temperature values remains and grows more prominent until $t_{\rm QSO} = 5 \times 10^7$ yrs. Between $t_{\rm QSO} = 5 \times 10^7$ yrs and $t_{\rm QSO} = 10^8$ yrs the temperature of the central bubble increases, leading to a broadening of the peak at the highest temperature values. As a result, the distinction between the two remaining peaks becomes less prominent.    

For reasons mentioned in the previous section, He is singly ionized along with H. However, for a bright QSO as the one considered here, the ionizing radiation is energetic enough to doubly ionize He appreciably, leading to the formation of a ring of HeII around the QSO (see K17). In the central panel of Figure \ref{fig:PDF-QSO} we show the distribution of $x_{\rm HeII}$. Unlike the cosmic reionization  simulations discussed earlier in which the fraction of cells with $x_{\rm HeII}\sim 1$ tends to unity, this is not observed here, as a substantial fraction of cells is being converted to doubly ionized helium. For the same reason, the peak at intermediate values of the ionization fraction is not as prominent as those seen in Figure~\ref{fig:PDF-GXQI} (central panel).

\begin{figure}
	\includegraphics[width=\columnwidth]{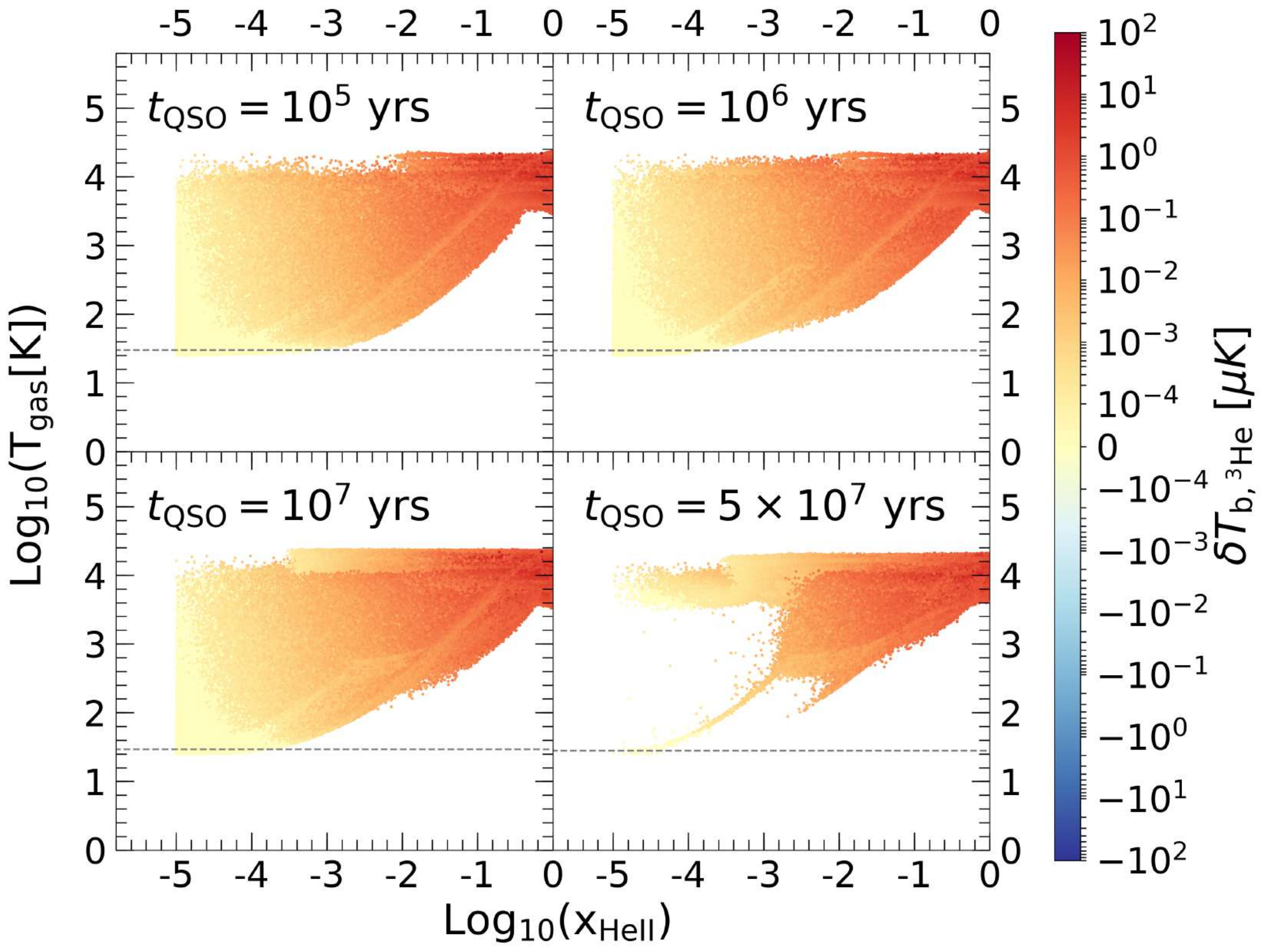}
   \caption{Differential brightness temperature, $\delta T_{\rm b,^3He}$, for the 256$^3$ cells in the simulation of a high-$z$ QSO as described in \cref{sec:QSO} as a function of the corresponding $T_{\rm gas}$ and $x_{\rm HeII}$ values. The panels refer to $t_{\rm QSO}=10^5$~yrs (top left), $10^7$~yrs (bottom left), $5 \times 10^7$~yrs (bottom right). The horizontal lines indicate $T_{\rm CMB}$.}
    \label{fig:DBT-Temp-xHeII-QSO}
\end{figure}

The corresponding distribution of the differential brightness temperature field is shown in Figure \ref{fig:PDF-QSO} (right panel). Due to the small amount of singly ionized He, initially $\delta T_{\rm b,^3He}$  is mostly zero. However, unlike in the cosmic reionization simulation, the signal can only be observed in emission with virtually nothing in absorption. The distribution resembles that of the corresponding $x_{\rm HeII}$ values, with the distinction between the peaks becoming less prominent with increasing QSO lifetime. The signal does, however, decrease slightly in intensity as redshift decreases, following the behaviour of the gas temperature.

\begin{figure}
	\includegraphics[width=\columnwidth]{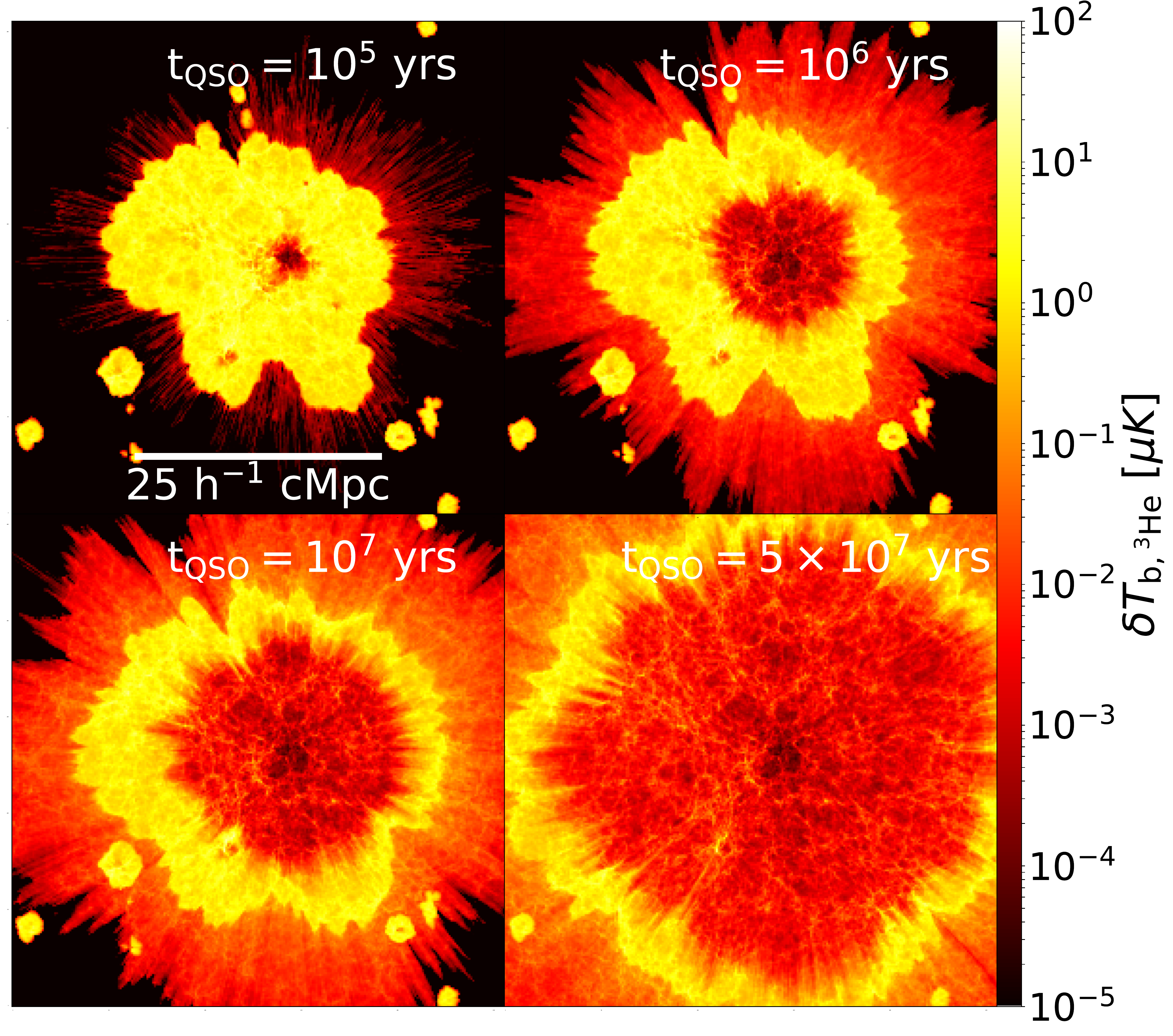}
   \caption{A slice of the differential brightness temperature field (in $\mu$K) in the simulation of a high-$z$ QSO as described in \cref{sec:QSO} at $t_{\rm QSO}=10^5$~yrs (top left), $10^6$~yrs (top right), $10^7$~yrs (bottom left), and $5 \times 10^7$~yrs (bottom right). }
    \label{fig:DBT-Field-QSO}
\end{figure}

In Figure \ref{fig:DBT-Temp-xHeII-QSO}, we plot a phase diagram illustrating the dependence of the differential brightness temperature on $T_{\rm gas}$ and $x_{\rm HeII}$. As the QSO gets older, the predominant contribution to the differential brightness temperature comes from highly ionized hot gas and the signal cannot be seen in absorption. The higher brightness of the QSO compared to that of the various BHs in the reionization simulations manifests itself also through the presence of hot gas with low $x_{\rm HeII}$. 

In Figure \ref{fig:DBT-Field-QSO} we show maps of $\delta T_{\rm b,^3He}$ in a slice of the simulation box. During the initial stages in the evolution of the QSO the ionized region surrounding it is mostly made of HeII, which can be seen as an almost homogeneous bubble with $x_{\rm HeII}\sim 1$ and $\delta T_{\rm b,^3He}$ as high as $\sim 50 \ \mu $K. 
However, as the QSO age increases, a region of doubly ionized helium rapidly forms and expands within the HeII bubble, leaving only a ring of outward expanding singly ionized helium, the thickness of which decreases with time. The peak of the emission arises from this ring, reaching a maximum of $\delta T_{\rm b,^3He} \sim $ 50 $\ \mu $K. The more energetic photons emitted by the QSO, due to their longer mean free path, are able to propagate beyond this region and to partially ionize helium, giving rise to a signal of $ \sim 10^{-3} \mu $K. Finally, the internal region, in which most helium is in its doubly ionized state, results in a differential brightness temperature between $\sim 10^{-3}$ and $10^{-2} \ \mu$K.  

\begin{figure}
	\includegraphics[width=\columnwidth]{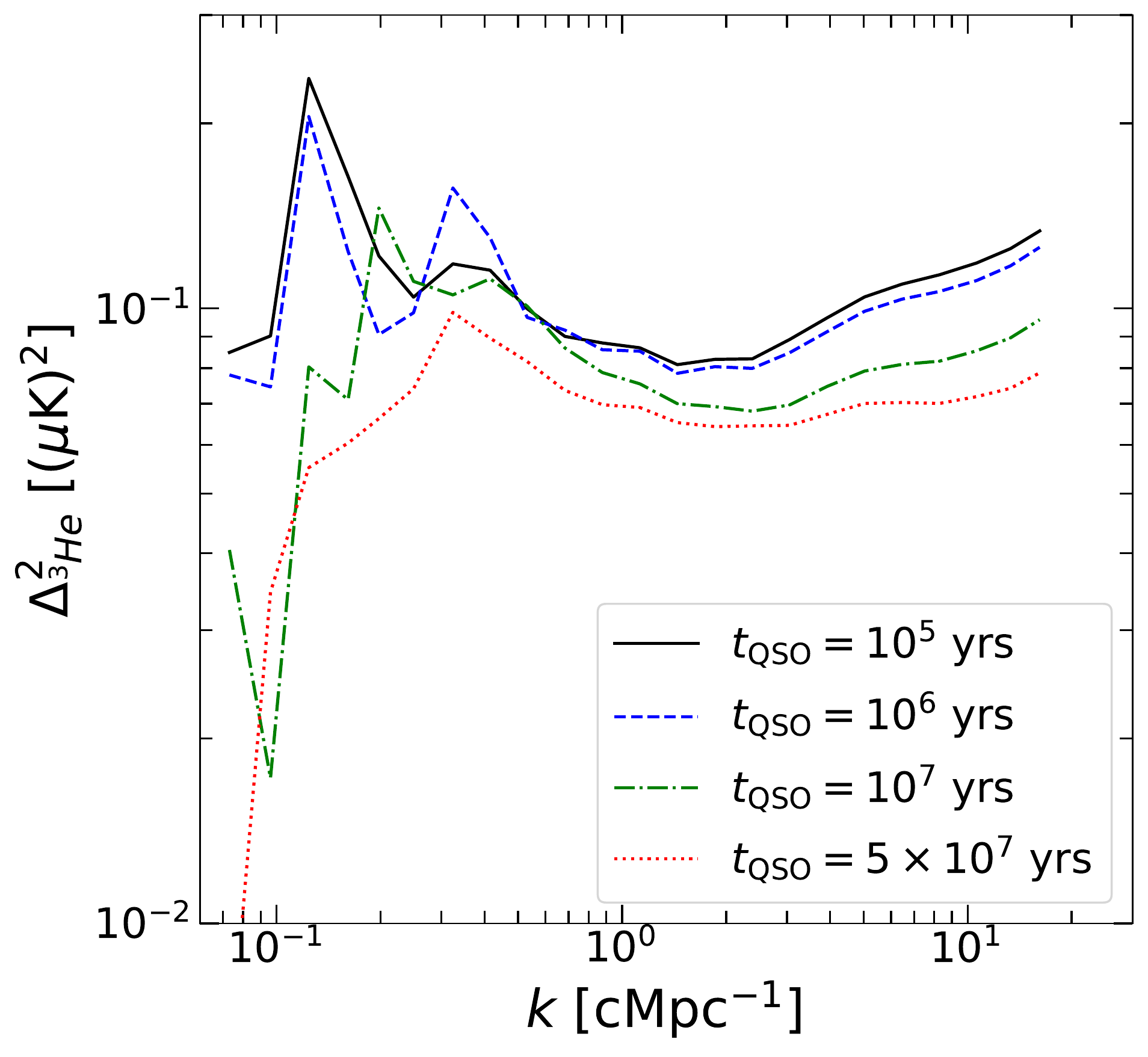}
   \caption{Power spectrum of the $^3$He$^+$ signal from the QSO simulations corresponding to QSO lifetimes of $t_{\rm QSO}= 10^5$~yrs (black solid line), $10^6$~yrs (blue dashed), $10^7$~yrs (green dash-dotted), and $5 \times 10^7$~yrs (red dotted). }
    \label{fig:PS-He-QSO}
\end{figure}

In Figure \ref{fig:PS-He-QSO} we show the power spectrum of the differential brightness temperature. The shape of the PS remains similar throughout the QSO's evolution, although it becomes flatter with time, i.e. the power is concentrated at the smallest and largest scales encompassed by the simulations, with an inflection at $k \sim 1$~cMpc$^{-1}$. As the QSO gets older, the HeII region surrounding it grows in size while a smaller HeIII region is formed closer to the QSO, resulting in an increase of power on large scales and a decrease on small scales. Eventually, though, the power decreases on all scales as less HeII is present.
The overall magnitude of the intensity of the PS is similar to the one obtained from the simulations of cosmic reionization described in \cref{sec:simul}.  

\subsection{Detectability}

In this section, we analyze the prospects of detecting the $^3$He$^+$ signal, which corresponds to a frequency of 0.58-1.24~GHz in the redshift range $z=6-14$.
Thus we take SKA1-mid\footnote{https://www.skatelescope.org/wp-content/uploads/2014/11/SKA-TEL-SKO-0000002-AG-BD-DD-Rev01-SKA1\_System\_Baseline\_Design.pdf} as the reference telescope. SKA1-mid is designed to cover the frequency range $0.35 - 14$ GHz, with a total of 190 15m dishes and 64 MeerKAT dishes, i.e. $N_{\rm dish} = 254$. It is expected to have a sensitivity $S_{N} = 1600 \,\rm m^{2}/K$ in the frequency range $\nu_{0} = 0.95 - 1.76\,\rm GHz$ and $S_{N} = 992 \,\rm m^{2}/K$ in the frequency range $\nu_{0} = 0.35 - 1.05\,\rm GHz$.
To test the detectability of the HeII signal during the EoR, we assume all the dishes to be in a core of radius $R=4\,\rm km$. 
The flux noise of SKA1-mid can be written as \citep{Wilson2009}:
\begin{equation}
\label{eq:F-N}
    \sigma_{N} = \frac{2k_{\rm B}T_{\rm sys}}{A_{\rm eff}\sqrt{N_{\rm dish}(N_{\rm dish}-1) B t_{\rm int}}},
\end{equation}
where $k_{\rm B}$ is the Boltzmann constant, $T_{\rm sys}$ is the system temperature of the telescope, $A_{\rm eff}$ is the effective collecting area of one dish, $N_{\rm dish}$ is the number of dishes, $B$ is the frequency bandwith and $t_{\rm int}$ is the integration time. 
Using the Rayleigh-Jeans relation $\sigma_{N} = 2k_{\rm B} T_{N} \Gamma_{b} \lambda^{-2}$, and the definition of sensitivity of the telescope $S_{N} = N_{\rm dish}* A_{\rm eff}/ T_{\rm sys}$, the rms of the brightness temperature of the noise can be written as:
\begin{equation}
\label{eq:DBT-N}
    \delta T_{N} = \frac{\lambda^{2}}{S_N/N_{\rm dish}\Gamma_{b}\sqrt{N_{\rm dish}(N_{\rm dish}-1) B t_{\rm int}}},
\end{equation}
where $\lambda$ is the wavelength and $\Gamma_{b} = 1.33(\lambda/R)^{2}$.
Assuming white noise, its power spectrum is $N(k) = \delta T_{N}^{2}V_{N}$, where $V_{N}$ is  the volume covered by the noise pixel.
We assume $B=0.1\,\rm MHz$ and $t_{\rm int} = 3000\,\rm h$. 
The signal to noise ratio (S/N) of the power spectrum can be expressed as:
\begin{equation}
\label{eq:S-N}
    \left(\frac{\rm S}{\rm N}\right)^{2} = \frac{4\pi k^{2} k_{\rm width} V}{(2\pi)^{3}}\frac{ P(k)^{2} }{[P(k)+N(k)]^{2}},
\end{equation}
where  $\theta = \lambda/R$, $V$ is the volume of the survey, and $k_{\rm width}$ is the width of the $k$ bin.

\begin{figure}
	\includegraphics[width=\columnwidth]{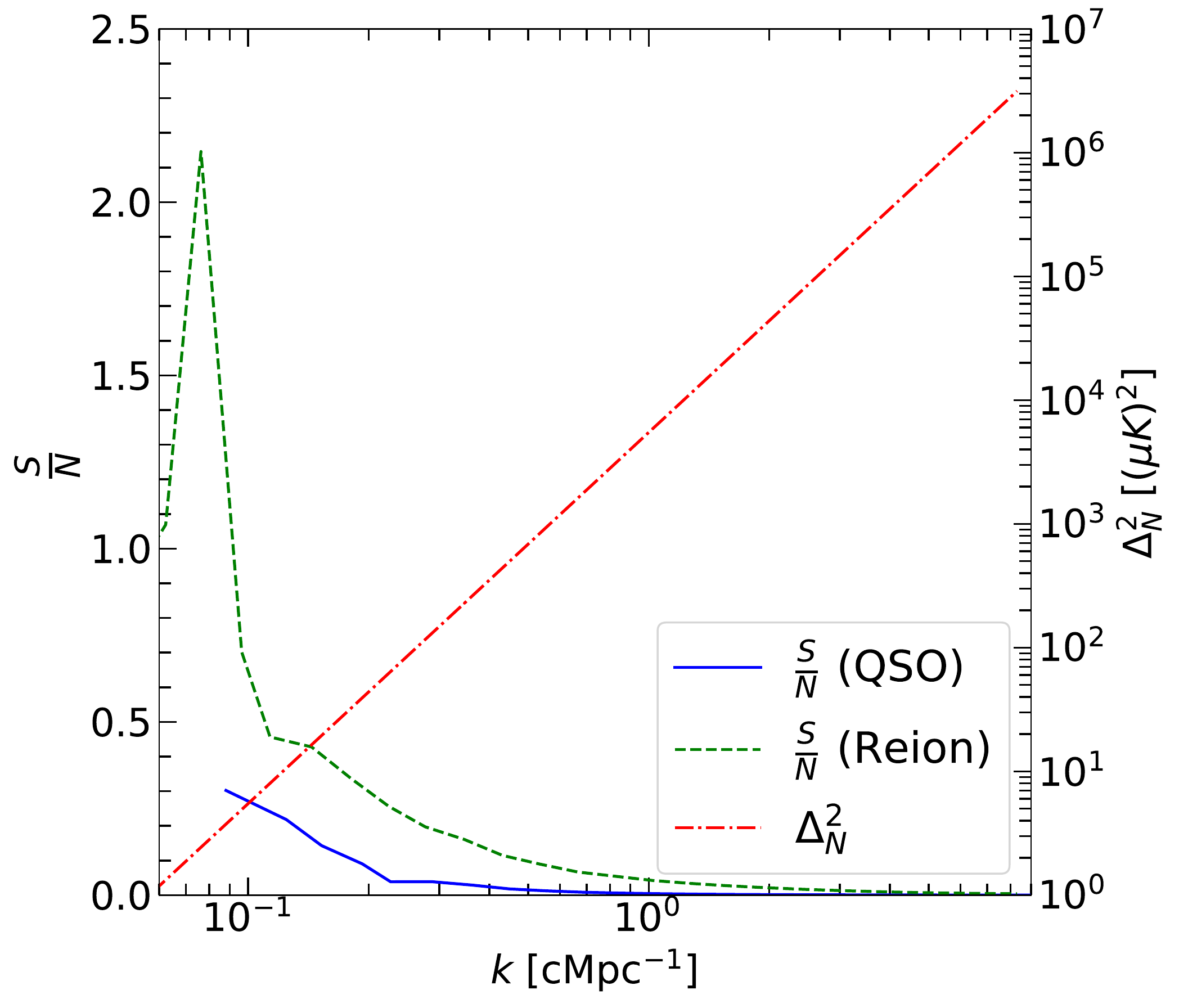}
   \caption{Signal to noise ratio (left y-axis) and the equivalent $\Delta^2_N$ (right y-axis) as a function of $k$. The green dashed and blue solid lines refer to the S/N of the $^3$He$^+$ signal (from eq.~\ref{eq:S-N}) in the cosmic reionization simulation at $z = 7$ and the high-$z$ QSO simulation with $t_{\rm QSO}=10^5$ yrs, respectively. The red dash-dotted line indicates the dimensionless power spectrum of the noise for our chosen reference telescope, SKA1-mid, with a frequency band $B=0.1\,\rm MHz$, an integration time $t_{\rm int} = 3000\,\rm h$, a $k$ bin width $k_{\rm width} = 0.23*k$ cMpc$^{-1}$ and a survey volume $V = 10^8$ cMpc$^3$.}
    \label{fig:Signal-noise}
\end{figure}

In Figure \ref{fig:Signal-noise}, we plot the quantity $\Delta_N^2$ using equations \ref{eq:Pk} and \ref{eq:Deltasquare} for the differential brightness temperature of the noise alongside the signal to noise ratios for the power spectra of the simulations of cosmic reionization (Figure \ref{fig:PS-He-GXQI}) at $z=7$. The S/N ratios are plotted using a reference $k$ bin width $k_{\rm width} = 0.23*k$ cMpc$^{-1}$  ( i.e. $\delta {\rm log}_{10}(k) = 0.1$),
and a survey volume $V = 10^8$ cMpc$^3$, which corresponds to a survey area $\Omega = 55.4 \ \rm deg^{2}$ at $z=7$, or equivalently, $\Omega = 51.5 \ \rm deg^{2}$ at $z=8$ with a slice width equal to $100 \ \rm cMpc$.
Comparing $\Delta_N^2$ to the PS of the signal plotted in Figs.~\ref{fig:PS-He-GXQI} and~\ref{fig:PS-He-QSO}, it is clear that the noise power spectrum on most scales is orders of magnitude higher than the corresponding values for the power spectrum of the signal. However, for large enough $k$ bin widths and survey volumes, the signal to noise ratio of a few can be reached up to $\sim 0.5$ cMpc$^{-1}$ for the cosmic reionization simulation. These scales are not covered by the QSO simulation (because of the smaller box), but also in this case we expect similar values.
Thus, our analysis does not rule out the possibility that telescopes in the future might be able to detect the signal.  

\section{Discussion and Conclusions}
\label{sec:Discussion and Conclusions}

In this paper we have evaluated the expected signal from the $^3$He$^+$ hyperfine transition using hydrodynamical and radiative transfer simulations of cosmic reionization which include different source types (stars, accreting nuclear BHs, XRBs and shock heated ISM; see E18 and E20), as well as of the environment of a bright QSO at $z=10$ (see K17).  
In both cases we find that the peak of the signal is expected to lie in the range $\sim$ 1-50 $\mu$K. While in the QSO's environment the signal is always in emission, in the case of cosmic reionization we observe a brief period in which the signal is expected to be also in absorption, with a maximum value of $\sim -10^{-3}$~$\mu$K. This is due to gas far away from sources, which gets partially ionized by energetic photons emitted by either the ISM or XRBs. 

We note that these results are valid under the assumption of coupling between the spin and gas temperature. Although a proper assessment of the coupling strength requires a detailed HeII Ly$\alpha$ radiative transfer, our approximate approach in Appendix~\ref{app} indicates that the HeII Ly$\alpha$ background produced in the cosmic reionization simulations is at least an order of magnitude below the one required for a full coupling through scattering. This suggests that, unless additional sources contributing to the background (such as a binary component in the stellar spectrum, a more abundant population of BHs at high redshift, the Ly$\alpha$ flux from excitation by X-ray photons) are included, an efficient coupling could be expected only in very high density pockets of gas through collisions, or in the vicinity of strong sources such as BHs (see also \citealt{Vasiliev2019}), where the local radiation dominates over the background (see e.g. \citealt{Ciardi2000}).

\begin{figure}
    \centering
    \includegraphics[width = \columnwidth]{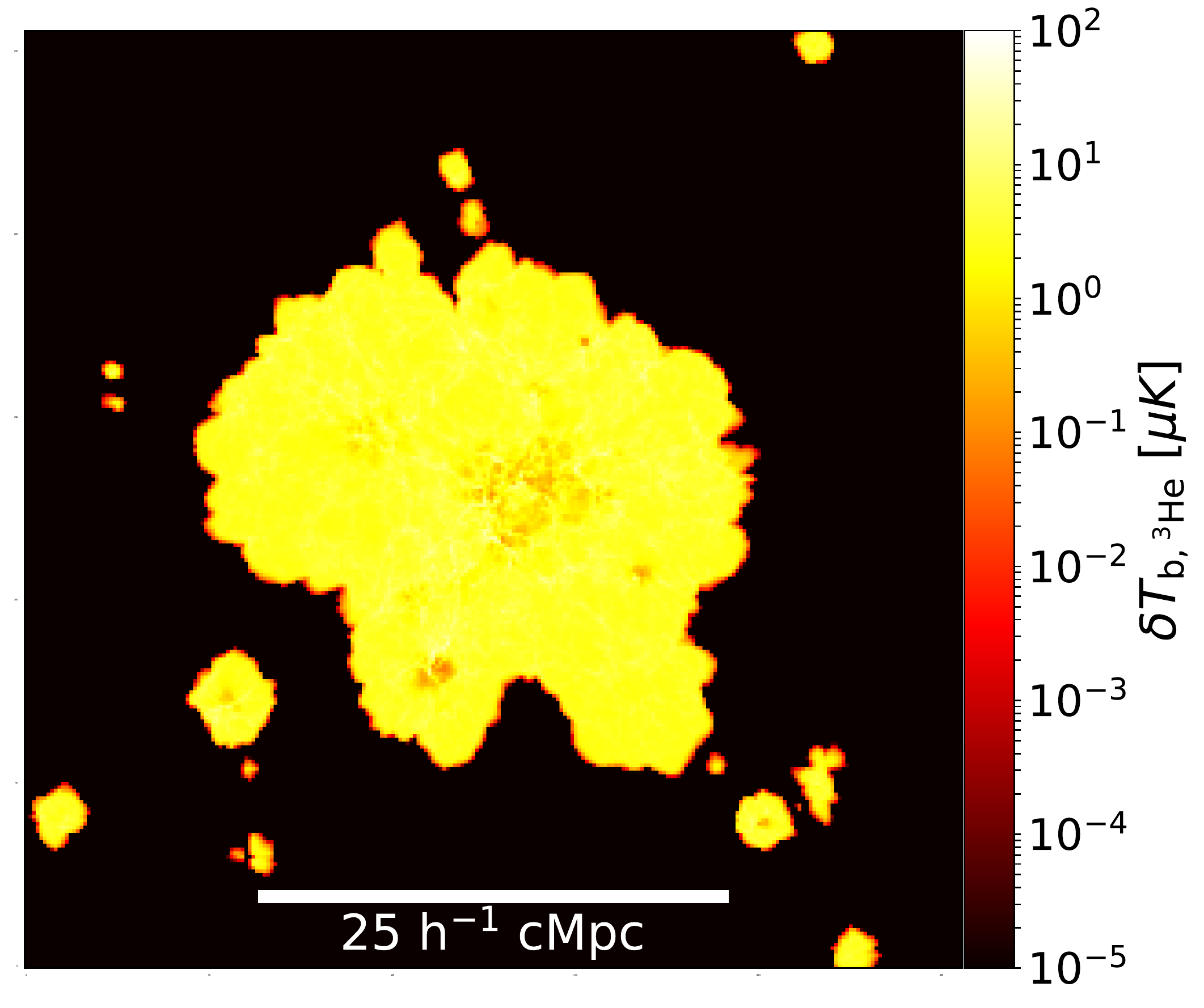}
    \caption{A slice of the $^3$He$^+$ differential brightness temperature field (in $\mu$K) in the simulation of the high-$z$ QSO described in \cref{sec:QSO}, but including only the galactic contribution to the ionizing radiation, i.e. the QSO is not turned on. The map is shown at a time corresponding to the bottom left panel of Fig.~7.}
    \label{fig:DBT-21cm-3He-comparison}
\end{figure}

It is important to note that it is not possible to distinguish reionization histories from different source types as we do not find any appreciable difference in the power spectrum between simulations containing only galactic stellar sources and the ones described in \ref{sec:simul} containing also XRBs, BHs and ISM. This happens because most of the signal is dominated by gas with $x_{\rm HeII}\sim 1$,  which in standard scenarios is driven by stellar type sources (see also discussions in E18 and E20). On the other hand, we do find that the $^3$He$^+$ signal might be a powerful probe to identify the presence of a bright high-$z$ QSO. In fact, the 21~cm signal associated to the environment of a QSO is very similar to the one from a large collection of galaxies (see Ma et al. submitted), because of the similarities in the associated ionized regions. On the contrary, the $^3$He$^+$ signal is very peculiar, as it can be seen by comparing the bottom left panel of Figure~\ref{fig:DBT-Field-QSO} to Figure~\ref{fig:DBT-21cm-3He-comparison}, where the differential brightness temperature is shown in the absence of the QSO contribution to reionization. In this case, the characteristic ``hole'' in the emission associated to the presence of HeIII is missing due to the softer spectrum of stellar type sources, which are not able to fully ionize He. 
This also suggests that, in principle, the 3.5~cm signal could be used to distinguish a QSO from a stellar-dominated reionization scenario, and potentially constrain their relative importance.

Finally, we find that the $^3$He$^+$ signal is weak and the prospects of its detection are presently severely limited. Using the SKA1-mid as our reference telescope, we find that the noise power spectrum dominates over the power spectrum of the $^3$He$^+$ signal. However, for a large enough $k$ bin width and survey volume, a signal to noise ratio of a few could be reached on the largest scales. Thus, our analysis leaves the possibility open for future telescopes to detect the $^3$He$^+$ hyperfine transition signal. 

\section*{Acknowledgements}

The authors would like to thank Aniket Bhagwat for helpful discussions, and Enrico Garaldi and an anonymous referee for their  insightful comments.
MBE is grateful the Institute of Theoretical Astrophysics at UiO for their kind hospitality.
QM is supported by the innovation and entrepreneurial project of Guizhou province for high-level overseas talents (grant no. (2019)02), the National Natural Science Foundation of China (grants No. 11903010), and the Science and Technology Fund of Guizhou Province (grants No. (2020)1Y020).

\section*{Data availability}

No new data were generated or analysed in support of this research.



\bibliographystyle{mnras}
\bibliography{main.bib}



\appendix
\section{Coupling to the gas temperature}
\label{app}

The spin temperature of $^3$He$^+$ is determined by the collisional process and HeII Ly$\alpha$ radiative transfer via the HeII analog of the Wouthuysen-Field effect \citep{BaglaLoeb2009,McQuinn.Switzer_2009,Takeuchi.Zaroubi.Sugiyama_2014}:
\begin{equation}
 T_{\rm s}=\frac{T_{\rm CMB}+y_{\rm c}T_{\rm k}+y_{\rm \alpha}T_{\rm \alpha}}{1+y_{\rm c}+y_{\rm \alpha}},
\end{equation}
where $y_{\rm c}$ and $y_{\rm \alpha}$ are the collisional and HeII Ly$\alpha$ coupling coefficients, respectively, $T_{\rm \alpha}$ is the color temperature of the radiation field near the $^3$He$^+$ Ly$\alpha$ line, and $T_{\rm k}$ is the kinetic temperature of the gas\footnote{Note that because we adopt the formulation of the spin temperature of \cite{Takeuchi.Zaroubi.Sugiyama_2014}, which is based on the original \cite{Field1958} paper, the collisional and Ly$\alpha$ coupling coefficients $y_c$ and $y_\alpha$ are not equivalent to the $x_c$ and $x_\alpha$ employed by \cite{Furlanetto2006} and \cite{McQuinn.Switzer_2009}.}.

\begin{figure}
    \centering
    \includegraphics[width = \columnwidth]{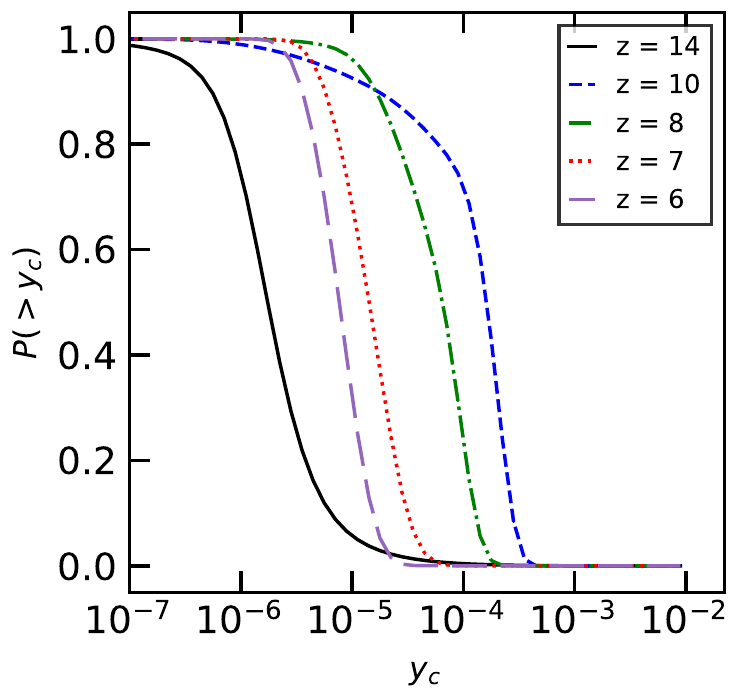}
    \caption{Probability for gas to have a collisional coupling coefficient larger than $y_{\rm c}$ for the simulation of cosmic reionization as described in \cref{sec:simul} at $z=14$ (black solid line), 10 (blue dashed), 8 (green dotted dashed), 7 (red dotted) and 6 (purple long dashed).}
    \label{fig:yc}
\end{figure}

\begin{figure}
    \centering
    \includegraphics[width = \columnwidth]{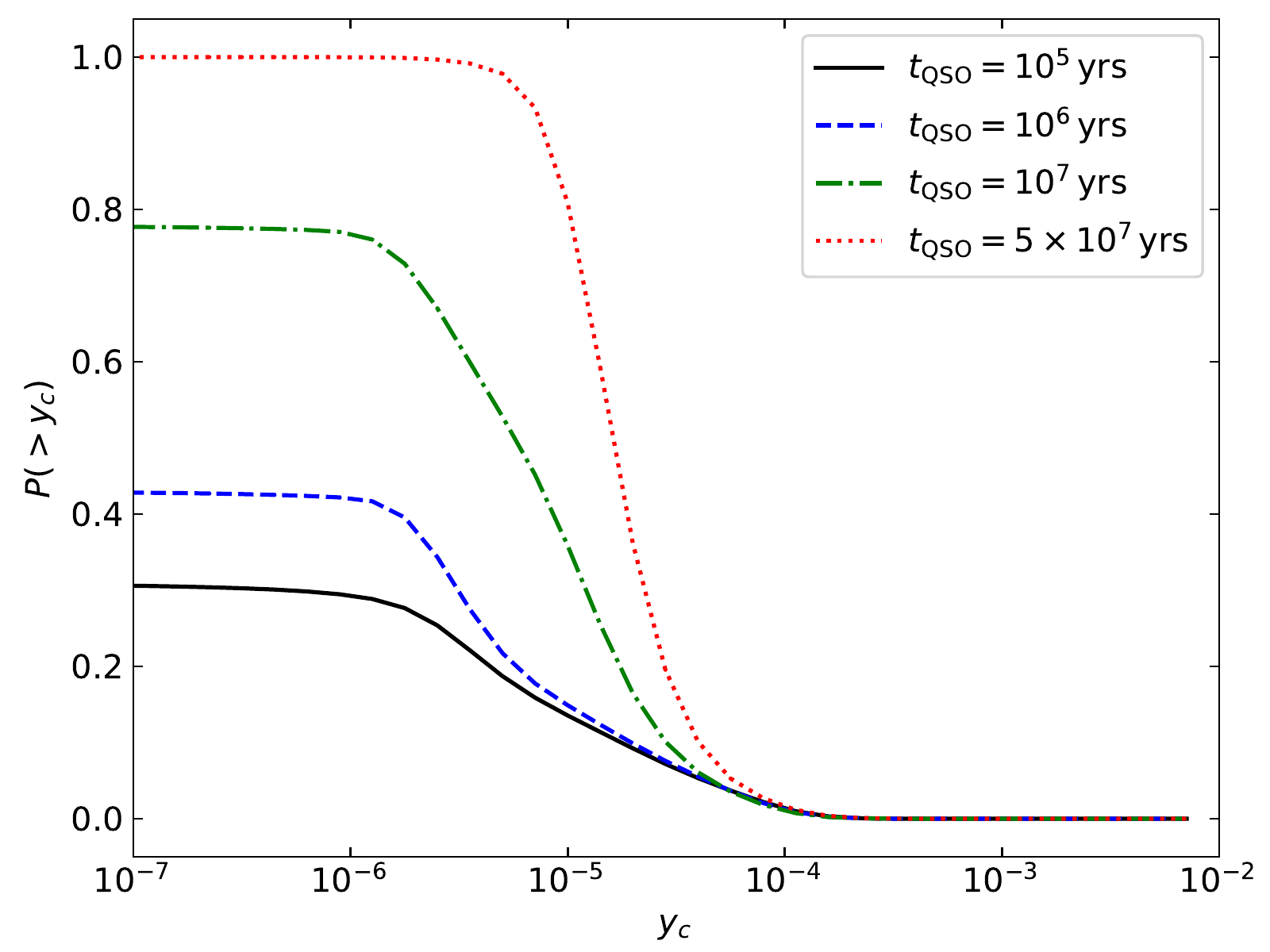}
    \caption{Probability for gas to have a collisional coupling coefficient larger than $y_{\rm c}$ for the QSO simulations corresponding to QSO lifetimes of $t_{\rm QSO}= 10^5$~yrs (black solid line), $10^6$~yrs (blue dashed), $10^7$~yrs (green dash-dotted), and $5 \times 10^7$~yrs (red dotted).}
    \label{fig:yc_qso}
\end{figure}

As for the 21~cm line from neutral hydrogen, collisional coupling is efficient exclusively at high density and low temperatures ($y_{\rm c} \propto n_{\rm e} \, T_{\rm k}^{-1.5}$, with $n_{\rm e}$ electron number density). In the typical conditions of the IGM, this is true only at $z \gtrsim 20$, as can be seen from Figure~\ref{fig:yc}, where no gas is found with $y_{\rm c}>10^{-3}$ at the redshifts of interest here. Similar results are found from the QSO's environment in Figure~\ref{fig:yc_qso}.
It should be noted that, employing simulations capable of resolving smaller scales, these would also capture pockets of higher density gas (e.g. Lyman limit systems), where the value of the collisional coefficient would be proportionally higher.
It should also be noted that a 21~cm signal would be produced also without full coupling, albeit with a smaller intensity.

The $^3$He$^+$ line can also be pumped through the HeII analog of the Wouthuysen-Field mechanism, i.e. by scattering off of UV photons at HeII Ly$\alpha$ line (40.8~eV). However, in this case the efficiency of the Wouthuysen-Field mechanism is less clear than for HI. Because of the presence of more abundant $^4$He$^+$ by a factor of $y_{^4{\rm He}}/y_{^3{\rm He}} \sim 7.5 \times 10^3$  ($y_{^3{\rm He}} \sim 10^{-5}$ and $y_{^4{\rm He}}=0.083$ are the primordial abundance ratio by number of helium isotope 3 and 4 relative to H atoms), most of HeII Ly$\alpha$ scatterings is caused by $^4$He$^+$. This led previous studies (e.g. \citealt{ChuzhoyShapiro2006,McQuinn.Switzer_2009}) to conclude that the Wouthuysen-Field mechanism is inefficient for $^3$He$^+$. However, this ignores the effect of multiple scatterings of HeII Ly$\alpha$ photons, which have a high HeII Gunn-Peterson optical depth of $\tau_{\rm GP}^{\rm HeII}=(y_{^4{\rm He}}/4)\tau_{\rm GP}^{\rm HII} \sim 1.6 \times 10^4 [(1+z)/11]^{3/2}$. This can effectively compensate the low abundance of $^3$He$^+$, because $(y_{^3{\rm He}}/y_{^4{\rm He}}) \tau_{\rm GP}^{\rm HeII} \sim 2$,  and thus boost the previous estimates of the HeII Wouthuysen-Field effect by a factor of 2.

Moreover, such investigations have considered the Wouthuysen-Field coupling strength by a homogeneous UV background. While this might be more relevant for a typical region of the diffuse IGM with stars only, in the presence of more energetic sources, these provide an additional reservoir of high energy UV photons blueward of the HeII Ly$\alpha$ line ($>$ 40.8~eV).
Therefore the Wouthuysen-Field coupling strength may not be as low as previously estimated, although a detailed modelling by solving the He II Ly$\alpha$ radiative transfer in an expanding universe is necessary to draw a quantitative conclusion, which is beyond the scope of the present investigation.

For a rough estimate of the HeII Ly$\alpha$ coupling strength, we follow \citet[][eqs. 2-4]{CiardiMadau2003} and evaluate the minimum Ly$\alpha$ background, $J^{\rm HeII}_{\rm th}$, required for the coupling to be effective, finding\footnote{The values of the various constants are taken from \cite{Takeuchi.Zaroubi.Sugiyama_2014}.}:
\begin{equation}
    J^{\rm HeII}_{\alpha}>J^{\rm HeII}_{\rm th}\approx 10^{-20} (1+z) \; {\rm ergs} \, {\rm cm}^{-2} \, {\rm s}^{-1} \, {\rm Hz}^{-1} \, {\rm sr}^{-1},
\end{equation}
which is about one order of magnitude higher than for efficient Ly$\alpha$ coupling for H.

\begin{figure}
    \centering
    \includegraphics[width = \columnwidth]{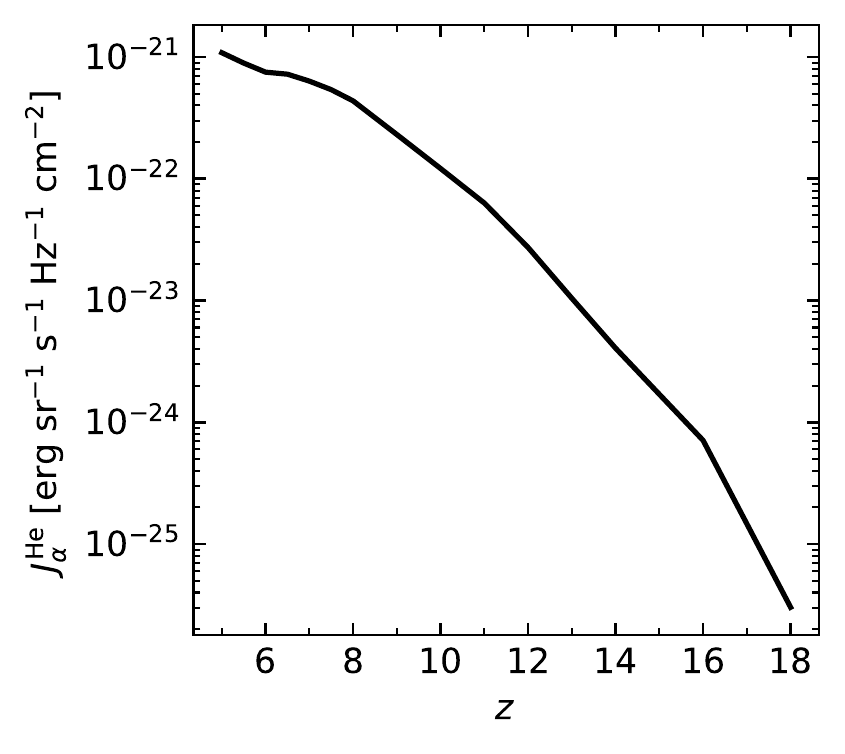}
    \caption{Evolution of the HeII Ly$\alpha$ background for the simulation of cosmic reionization.}
    \label{fig:Ja}
\end{figure}

In Figure~\ref{fig:Ja} we show the redshift evolution of the HeII Ly$\alpha$ background obtained from our simulations of cosmic reionization. We see that at any time the background is at least an order of magnitude lower than what required for a full coupling. The addition of the contribution from sources which have not been included in this work, such as a binary component in the stellar spectrum, a more abundant population of BHs at high redshift, the Ly$\alpha$ flux from excitation by X-ray photons (see e.g. \citealt{PritchardFurlanetto2007}), could raise the He Ly$\alpha$ background above $J^{\rm HeII}_{\rm th}$, but, as mentioned earlier, a proper quantitative assessment of the Ly$\alpha$ coupling requires a more accurate physical modeling. 

Note also that a full coupling of the color temperature of the HeII Ly$\alpha$ radiation field to the spin temperature is not necessary to observe the differential brightness temperature in emission, as long as the coupling rises the color temperature sufficiently above the CMB temperature (e.g. \citealt{DeguchiWatson1985}).

Finally, as the radiation from local sources is expected to dominate over the background radiation, in particular at the higher redshifts before a strong background had time to build up (see e.g. \citealt{Ciardi2000}), the best (possibly only) locations where the signal could be detected would be in the vicinity of the sources, in particular BHs (see also \citealt{Vasiliev2019}).


\bsp	
\label{lastpage}
\end{document}